\documentstyle[aps,prb,epsf,graphicx]{revtex}

\begin{document}

\title{Polaron formation for a non-local electron-phonon coupling: A variational wave-function study}
\author{C. A. Perroni,$^{1,2}$ E. Piegari,$^{2}$  M. Capone,$^{3}$ and V. Cataudella$^{1,2}$}
\address{$^{1}$Coherentia-INFM, UdR di Napoli, via Cinthia 80126 Napoli (Na), Italy  \\
$^{2}$Dipartimento di Fisica, Universit\'a di Napoli ``Federico II'', Italy \\
$^{3}$Enrico Fermi Center, Rome, Italy}

\date{\today}
\maketitle
\begin{abstract}
We introduce a variational wave-function to study the polaron
formation when the electronic transfer integral depends on the
relative displacement between nearest-neighbor sites giving rise
to a non-local electron-phonon coupling with optical phonon modes.
We characterize the polaron crossover by analysing 
ground state properties such as the energy, the
electron-lattice correlation function, the average phonon occupation and the
quasiparticle spectral weight. Variational results are found in good agreement
with numerical exact diagonalization of small clusters,and follow the 
correct perturbative result at weak coupling. We determine the
polaronic phase diagram and we find that the tendency towards strong
localization is hindered from the pathological sign change of the
effective next-nearest-neighbor hopping.
\end{abstract}

\section{Introduction}

Significant electron-phonon (el-ph) interactions have been experimentally 
detected in many materials of wide interest, like manganites, \cite{manganiti} 
fullerenes, \cite{fullerene}, carbon nanotubes, \cite{nanotubi,baughman} and 
cuprates. \cite{cuprati} In many of these cases, the el-ph interaction
gives rise to polaronic features.
A polaronic state results in fact when the electrons are strongly coupled
to  lattice distortion, therefore
increasing their effective mass and leading to a state with low
mobility. Increasing the el-ph coupling, the spatial extension of
the lattice deformation decreases \cite{alexandrov} and the
polaron can vary its size from {\em{large}} to {\em{small}}. 
The single polaron problem of one electron interacting with 
the lattice degrees of freedom has been studied in detail, and 
allowed us to understand in the detail the physics leading to 
the formation of polaronic states. In
particular it has been shown that the self-trapping process, 
which lead to the formation of polarons, is not a phase
transition, but just a continuous crossover with no broken
symmetry. \cite{gerlach} In the case of the Holstein model,
\cite{holstein} where quantum vibrations interact {\em locally}
with the electrons, the crossover from large to small polaron has
been extensively studied by several numerical techniques
\cite{7,korn,8,massimo,10,ciuchi,ccg} and variational approaches.
\cite{romero,trugman,cataud1}  In particular all the ground state
properties of the Holstein model can be described with great
accuracy by a variational approach \cite{cataud1} based on a
linear superposition of Bloch states that describe weak and strong
coupling polaron wave functions.

The case of non-local interactions, that in general are also
present in real materials, is much less understood. The coupling
with acoustical phonons 
has been studied in order to explain the anomalous transport
properties of non-local excitations, like solitons and polarons,
in various 1D systems. \cite{ssh,Lu,angilella,zoli} 
In particular the tight-binding Su-Schrieffer-Heeger (SSH) model \cite{ssh} 
was introduced to explain the transport properties of quasi one-dimensional
polymers  as polyacetylene where the CH monomers form chains of alternating 
double and single bonds. In this case the localization is due to a large 
shrink of two particular bonds and corresponding large hopping integral 
between the sites. As a result, the hopping between the two occupied sites
and the surrounding ones is reduced resulting in a tendency towards 
localization.
 

Our purpose here is to examine the single polaron
formation in a model where both (Holstein) local and (SSH) non-local
el-ph interactions are present. Due to the complexity of the
model, we start the analysis using a perturbative approach that,
although it can not capture the full multiphononic nature of the
polaron, it has proved a remarkably useful tool in understanding
the el-ph physics. \cite{massimo,zoli} In particular we
characterize the ground state properties of the system evaluating
the energy, the electron-lattice correlation function and the
quasiparticle spectral weight, in order to provide signatures of the polaron
formation. In the limit when local el-ph interactions are much
stronger or weaker than non-local el-ph interactions our model reduces
to the standard Holstein model and to the SSH model with a
dispersionless phonon spectrum, respectively. Then we start an
accurate analysis of the non-local limit case, being this case not
yet fully examined. Recently, the fully adiabatic regime of this
model has been used to explain changes in carbon-nanotube length
as a function of charge injection. \cite{baughman} Furthermore the
non-local case has been previously studied by one of us using
exact diagonalization of small clusters up to four lattice sites, 
where an anomalous optical absorption has been identified. \cite{massimo} 
In particular in Ref. [\onlinecite{massimo}] it is shown that the
strong-coupling solution is characterized by an unphysical sign
change of the effective next-nearest-neighbor hopping which is
missing when acoustical phonons are considered. \cite{angilella}

In this work we improve the previous numerical analysis
considering a six-site lattice and introduce a variational
wave-function to investigate the thermodynamic limit of the
system. The variational approach is based on a linear
superposition of Bloch states that provide an excellent
description of the lattice deformations on left and right bond of
the polaron, respectively. The wave-function closely resembles a
variational state previously proposed for the study of the
Holstein model \cite{cataud1} and for the SSH case it allows to
describe polaron features in good agreement with exact numerical
diagonalization results. The variational approach  recovers the
pathological behavior of the effective next-nearest-neighbor
hopping pointing out that the non-physical region of parameters
always prevents a strong localized solution. We also
explicitly show that, when the phonon frequency is not really
small, the considered non-local SSH interaction supplies a
tendency to localize for the single carrier which can be more
effective than the Holstein localization.

The scheme of the paper is the following. In Sec. II we present
the model and set down the notation. In Sec. III we discuss
perturbative calculations showing the role of SSH el-ph
coupling with respect to the Holstein contribution. Sec. IV is
devoted to the presentation of the variational method in the limit
of non-local el-ph interactions and its comparison with the exact
diagonalization results. Sec. V reports our concluding remarks.

\section{The Model}

In extremely general terms, the interaction between electron and
harmonic lattice deformations is described by the Hamiltonian
\begin{equation}
{\cal H} = \sum_{i,j,\sigma}c^{\dag}_{i,\sigma}t_{i,j}(\lbrace x_k
\rbrace ) c_{j,\sigma} + \sum_{i}\frac{p_i^2}{2M} +
\sum_{i,j}\frac{x_i K_{i,j}x_j}{2} + \sum_{i,\sigma} e_i (\lbrace
x_k \rbrace ) c^{\dag}_{i,\sigma} c_{i,\sigma},
 \label{hami}
\end{equation}
where $c^{\dag}_{i,\sigma}$ ($c_{i,\sigma}$) is the fermion
creation (destruction) operator, $\sigma$ is the spin index,
$t_{i,j}(\lbrace x_k \rbrace ) $ is the electronic transfer
integral for fixed lattice deformations $\lbrace x_k \rbrace$, $M$
is the ionic mass, $K_{i,j}$ is the spring constant matrix, and
$e_i (\lbrace x_k \rbrace )$ is the local energy of the electron.
For small deviations from the equilibrium positions of the lattice
we can approximate $t_{i,j}(\lbrace x_k \rbrace)$ and $e_i
(\lbrace x_k \rbrace )$ to be linear functions of the lattice
displacements $\lbrace x_k \rbrace$ obtaining a general model with
el-ph interactions. In particular, limiting the hopping to
nearest-neighbor sites of a linear chain, we make the assumption
\begin{equation}
t_{n+1,n}\left( \{x_k \} \right) = -t + \alpha (x_{n+1}-x_{n})
\label{hop}
\end{equation}
typically employed for the derivation of the el-ph SSH interaction
term, and
\begin{equation}
e_i (\lbrace x_k \rbrace )=\alpha_1 x_i
 \label{energloc}
\end{equation}
generally used in order to deduce the local el-ph Holstein
interaction. If spinless electrons and dispersionless Einstein
phonons are considered, the model becomes
\begin{equation}
{\cal H}= -t \sum_i (c^{\dag}_i c_{i+1}+c^{\dag}_{i+1} c_{i}) +
\omega_0 \sum_i a^{\dag}_{i} a_{i}+H_{int}, \label{hami1}
\end{equation}
where $H_{int}$ is

\begin{equation}
H_{int}= g\omega_0 \sum_i (c^{\dag}_i c_{i+1}+c^{\dag}_{i+1}
c_{i}) (a^{\dag}_{i+1} + a_{i+1} -a^{\dag}_{i}- a_{i}) +
g_1\omega_0 \sum_i c^{\dag}_i c_{i}(a^{\dag}_i+ a_{i}),
\end{equation}
with $a^{\dag}_{i}$ ($a_{i}$) the phonon creation (destruction)
operator and $\omega_0$ the quantum of vibrational energy per
site. The quantity $g=\alpha/\sqrt{2 M \omega_0^3}$ is the SSH
coupling that we mainly discuss in this work, while
$g_1=\alpha_1/\sqrt{2 M \omega_0^3}$ is the Holstein local
electron-phonon coupling. We study the coupling of a single
electron to lattice deformations using units such that the lattice
spacing $a=1$ and $\hbar =1$.

\section{Perturbation Theory}

Weak-coupling perturbation theory in the electron-phonon coupling
has proved a remarkably useful tool in understanding the el-ph
physics. Besides the obvious ability to describe the weakly
interacting regime, the perturbative approach has in fact provided
some guidelines to understand the conditions for polaron formation
in the Holstein model. More explicitly, the polaron crossover
occurs around the coupling value for which the perturbative
approach breaks down. \cite{massimo}

Here we focus on the case of one electron in a one-dimensional
chain. If the el-ph terms are smaller than both the hopping term
and the bare phonon term $(g, g_1 \ll \tilde t,1$ and $\tilde t
=t/\omega_0$), they can be treated as perturbations of the
unperturbed Hamiltonian $H_0=H_{kin}+H_{ph}$.

The  second-order correction to the energy of the ground state
is given by

\begin{equation}
\Delta E(0) = - g^2 \left(\frac{1+2\tilde t-\sqrt{1+4\tilde
t}}{\tilde t^2}\right) -g_1^2 \frac{1}{\sqrt{1+4\tilde t}},
\label{egs}
\end{equation}
while the perturbative correction to the free
band $\varepsilon_k = -2t\cos (k)$ is reported in Appendix A.

We note that for fixed values of the coupling constants $g$ and
$g_1$ the two  contributions (SSH-like and Holstein) have different
behaviors as functions of the inverse adiabatic ratio $\tilde t$.
In particular, as shown in Fig. 1, the Holstein contribution to
the ground state energy is always lower (for $g=g_1$) than the SSH
one when $\tilde t < \tilde t_w$ with $\tilde t_w=4+3\sqrt2$. In
other words, when the phonon frequency are not really small, the
SSH el-ph coupling is more effective than the Holstein one.
The reduced effect of the Holstein el-ph coupling when $\tilde t$ is 
small (anti-adiabatic regime) pushes the polaron crossover to 
larger values of the coupling $\lambda=g^2/2\omega_0t$ as the phonon frequency 
is increased. Actually, while $\lambda > 1$ is the condition
for the polaron crossover in the adiabatic regime $\tilde t \gg 1$,
in the antiadibatic regime $\tilde t \ll 1$,
it has been shown that the crossover occurs when $\alpha^2 =
g^2/\omega_0^2 \simeq 1$, i.e., for $\lambda \gg 1$. 
\cite{massimo,ciuchi,ccg} Recently it has been
shown that this important role of the degree of adiabaticity is not 
limited to the single polaron problem, but it also extends to 
finite dennsities. \cite{caponeciuk}

Since polaron formation is not a phase transition and occurs
without symmetry breaking, different criteria can be established
to define the crossover values of the coupling constants which
mark the polaronic regime. In the following we compute some
physical quantities which have been often introduced to
characterize the polaron crossover.

The average phonon occupation number $N_{ph}  =\frac{1}{N}\langle
\sum_i a^{\dagger}_i a_i\rangle$ is given by
\begin{equation}
N_{ph} =  \frac{2\lambda}{\tilde t}\left[
\frac{(1+2\tilde t) }{\sqrt{1 +4\tilde t }}-1\right]
+ 2\lambda_1 \tilde t \frac{(1+2\tilde t) }{(1 +4\tilde t)^{ 3/2}},
\label{nph0}
\end{equation}
where $\lambda = g^2 \omega_0 /2t$ and $\lambda_1 = g^2_1 \omega_0
/2t$. From Eq. (\ref{nph0}) it turns out that the phonon number, as
the ground state energy, is more affected by the SSH coupling when
$\omega_0$ exceeds a given value. In particular for $\tilde t \leq
2$ (i.e. $\omega_0 /t \geq 0.5$) the SSH contribution is always
higher than the Holstein one.

Other quantities of great interest to characterize the polaron
formation are the electron-lattice correlation functions. In
particular we consider the correlation function
$\chi_{i,\delta}=\langle c^{\dag}_i c_i (a^{\dag}_{i+\delta} + a_{i
+\delta})\rangle$ between the electronic density on a site $i$ and the
lattice displacement on site $i+\delta$, which measures the
entanglement of lattice and electronic degrees of freedom typical
of the polaronic state. After a Fourier transformation in the
momentum space, at $k=0$ one has
\begin{eqnarray}
\chi_{k=0,\delta =0} &=&  -\frac{2g_1}{\sqrt{1+4\tilde t }} \nonumber \\
&& \nonumber \\
\chi_{k=0,\delta =1} &=&  -\frac{g}{\tilde t^2}(1 + 2\tilde t - \sqrt{1
+4\tilde t }) -\frac{g_1}{\tilde t} \left(
\frac{2 \tilde t +1 }{\sqrt{1+4\tilde t }}
  -1 \right) \nonumber \\
&& \nonumber \\
\chi_{k=0,\delta =2} &=&  -2g \left[\frac{1}{\tilde t}+
\frac{1+4 \tilde t}{2\tilde t^3} \left(1-
\frac{1 + 2 \tilde t  }{\sqrt{1+4\tilde t }} \right)\right] -2g_1
\left[\frac{1}{\sqrt{1+4\tilde t }}+
\frac{1+2 \tilde t}{2\tilde t^2} \left(1-
\frac{1 + 2 \tilde t  }{\sqrt{1+4\tilde t }} \right) \right].
\label{F0}
\end{eqnarray}
In Fig. 2 we plot the correlation function at nearest-neighbor
(left) and next-nearest-neighbor (right) sites as functions of the
inverse adiabatic ratio $\tilde t$, for fixed values of the
couplings $g=g_1=1$. As expected, the value of the correlation
function goes to zero for large values of $\tilde t$, but the
behavior of the SSH-like contribution (dashed lines) is
qualitatively different from that of the Holstein ones (dot-dashed
lines).

The last quantity we consider is the quasiparticle spectral weight
$Z(k)= (1-\frac{\partial \Sigma(k,\omega)}
{\partial\omega}|_{\omega=\varepsilon(k)})^{-1}$, which measures
the renormalization of the electron Green's function due to the 
el-ph interaction. The second-order perturbative self-energy
$\Sigma(k,\omega)$ is given in Appendix A. Even if the polaronic
regime cannot be attained within lowest-order perturbative
approach, indications on the beginning of the polaronic crossover
can be extracted from the spectral weight expression. In
particular the polaron crossover is expected to be associated with
a sharp reduction of this quantity as a function of the couplings.
The expression of the inverse spectral weight at $k=0$ is given by
\begin{equation}
Z(0)^{-1} =  1-\frac{2\lambda}{\tilde t} \left[ 1-
\frac{(1+2\tilde t) }{\sqrt{1 +4\tilde t }} \right]
 + 2\lambda_1 \tilde t
\frac{(1+2\tilde t) }{(1 +4\tilde t)^{ 3/2}}, \label{zk0}
\end{equation}
while the full momentum dependence of $Z(k)$ is reported in
Appendix A. As expected the spectral weight $Z(0)$ is a monotonically
increasing function of $\tilde t$, for a fixed value of the
couplings. It is interesting to note that the reduction of $Z(0)$
due to the SSH-like contribution is more relevant of the Holstein
ones for $\tilde t < 2$, while in the adiabatic limit, i.e. for
large value of $\tilde t$, it is very small and slow.

Strictly speaking, perturbative calculations only correctly characterize
the small coupling regime. In order to provide a better insight on the
problem of the polaron formation in the systems with non-local interactions,
in the following we focus on the SSH contribution and substantiate our
results by analytic variational calculations and numerical exact data.

\section{Variational Approach vs. Exact Diagonalization}

In this section we extend our analysis of the non-local SSH model
to the whole range of el-ph couplings using two standard and well
grounded techniques, a variational approach and exact
diagonalization of small clusters. First we introduce the
variational wave function. We consider translation-invariant Bloch
states obtained by superposition of localized states
centered on different lattice sites. \cite{toyo} These
wave-functions have been introduced in order to study the polaron
formation within the Holstein model where they are able to fully
capture the features of the Holstein polaron.
\cite{romero,cataud1} In this work we extend this kind of
wave-functions to the SSH interaction model assuming
\begin{equation}
|\psi_k^{(i)} \rangle  =    \frac{1}{\sqrt{N}} \sum_n e^{ik \cdot
n} |\psi_k^{(i)}(n) \rangle, \label{psi}
\end{equation}
where $|\psi_k^{(i)}(n) \rangle$ is defined as
\begin{equation}
|\psi_k^{(i)}(n) \rangle =
e^{[U_k^{(i)}(n)+U_k^{(i)}(n-1)+U_k^{(i)}(n+1)]}
|0\rangle_{ph}\sum_m \phi_k^{(i)} (m)e^{ik \cdot m} c^{\dag}_{n+m}
|0\rangle_{el}, \label{psi1}
\end{equation}
with the quantity $U_k^{(i)}(j)$ given by
\begin{equation}
U_k^{(i)}(j) =  \frac{g}{\sqrt{N}}\sum_q [f^{(i)}_{k,j}(q) a_q
e^{iq \cdot R_j}-h.c.]. \label{trasformazioni}
\end{equation}
The phonon distribution function $f^{(i)}_{k,j}(q)$ is chosen as
\begin{equation}
f^{(i)}_{k,j}(q)  =  \frac{\alpha^{(i)}_{k,j}}
 {1+2\tilde t \beta^{(i)}_{k,j} \left[ cos(k)-\cos(k+q) \right]},
\label{distribuzione}
\end{equation}
with $\alpha^{(i)}_{k,j}$ and $\beta^{(i)}_{k,j}$ variational
parameters. In Eq. (\ref{psi1}), $|0\rangle_{ph}$ and
$|0\rangle_{el}$ denote the phonon and electron vacuum state,
respectively, and the variational functions $\phi_k^{(i)} (m)$ are
assumed to be
\begin{equation}
\phi_k^{(i)} (m)=\sum_{j=-5}^{5} \gamma_k^{(i)}(j) \delta_{m,j},
\label{extension}
\end{equation}
where $\gamma_k^{(i)}(j)$ are variational parameters that take
into account the broadening of the electron wave-function up to
fifth neighbors. It is worth to note that traditional variational
approaches to the Holstein polaron problem uses the localized
state (\ref{psi1}) where only the on-site operator $U_k^{(i)}(n)$
is applied.  Thus we introduce in the expression of the trial
wave-function the nearest-neighbor displacement operators
$U_k^{(i)}(n+1)$ and $U_k^{(i)}(n-1)$, in order to take into
account the dependence of the hopping integral on the relative
distance between two adjacent ions.

\noindent 
Reflecting the asymmetry of the SSH coupling
(shrinking of the bond on which the electron is localized
and stretching of the neighboring bonds), we also define
two wave-functions that provide a very good description of the
lattice deformations on left and right bonds of the polaron.
Naturally the left and right directions
are relative to the site where the presence of the electron is
more probable. Thus in Eq. (\ref{psi1}) the apex $i=L,R$ indicates
the $Left$ $(L)$ and $Right$ $(R)$ polaron wave-function,
respectively. The wave-functions $L$ and $R$ are related as
follows
\begin{eqnarray}
&& f_{k,n}^{(R)}(q)=-f_{k,n}^{(L)}(q)<0 \nonumber \\
&& f_{k,n-1}^{(R)}(q)=-f_{k,n-1}^{(L)}(q)>0 \nonumber \\
&& f_{k,n+1}^{(R)}(q)=-f_{k,n+1}^{(L)}(q)>0 \nonumber \\
&& \phi^{(R)}_k(m)=\phi^{(L)}_k (-m).
\label{degeneracy}
\end{eqnarray}

\noindent
All the variational parameters are determined by minimizing the
expectation value of the Hamiltonian (\ref{hami1}) with $g_1=0$ on
the states (\ref{psi1}). Even though the wave-functions $L$ and $R$
describe correctly the lattice deformations of the left and right
side of the polaron, respectively, the mean values of the
Hamiltonian on these states are equal. So the relations
(\ref{degeneracy}) can be also viewed as those that leave
unchanged the energy functional determined by one wave-function.

These two wave-functions can be improved by increasing the
extension of the phonon contributions in Eq. (\ref{psi1}) and of
the electron terms in Eq. (\ref{extension}). Furthermore, they are
not orthogonal and the off-diagonal matrix elements of the
Hamiltonian between these two states are not zero. This allows to
determine the ground-state energy by considering as trial state
the linear superposition \cite{cataud1} of the wave-functions $R$
and $L$
\begin{equation}
|\psi_k \rangle =  \frac{A_k|\Phi_k^{(R)} \rangle +
B_k|\Phi_k^{(L)} \rangle} {\sqrt{A_k^2 + B_k^2 + 2A_k B_k S_k}},
\label{psitot}
\end{equation}
where $|\Phi_k^{(L)} \rangle$ and $|\Phi_k^{(R)} \rangle$ are the
normalized wave-functions $L$ and $R$ weighted by the coefficients
$A_k$ and $B_k$ and
\begin{equation}
S_k =  \langle \Phi_k^{(L)} |\Phi_k^{(R)}\rangle
\end{equation}
is the overlap factor. The wave-function (\ref{psitot}) correctly
describes the properties of the lattice deformations on both the
sides of the polaron and we will find that it is in very good
agreement with the results derived by the exact diagonalizations
on a chain of 6 sites. Furthermore the variational approach
involves a number of variational parameters that does not depend
on the length of chain, so it allows to study the thermodynamic
limit of the system.

The minimization procedure is performed in two steps. First the
energies of the left and right wave-functions are separately
minimized, then these wave-functions are used in the minimization
procedure of the quantity $E_k=\langle \psi_k | H | \psi_k \rangle
/ \langle \psi_k |\psi_k\rangle$ with respect to $A_k$ and $B_k$
defined in (\ref{psitot}).
\cite{cataud1}
Exploiting the equality
\begin{equation}
\langle\psi_k^{(L)} |H|\psi_k^{(L)}\rangle
= \langle\psi_k^{(R)} |H|\psi_k^{(R)}\rangle \nonumber \\
= \varepsilon_k,
\end{equation}
we obtain
\begin{equation}
E_k =  \frac{\varepsilon_k - S_k E_{kc} - |E_{kc}-S_k
\varepsilon_k|}{1-S_k^2},
\label{enetot}
\end{equation}
where $E_{kc}=\langle\Phi_k^{(L)} |H|\Phi_k^{(R)}\rangle$ is the
off-diagonal matrix element, and $|A_k|=|B_k|$. The matrix
elements between the states $\psi_k^{(R)}$ and $\psi_k^{(L)}$
contained in Eq. (\ref{enetot}) are reported in Appendix B.

The total energy functional (\ref{enetot}) is minimized with
respect to the variational parameters and the optimal ground state
energy is plotted in Fig. 3 for a six-site lattice and two
different values of the inverse adiabatic parameter $\tilde t$. We
also study the thermodynamic limit and find energy curves very
close to those of the finite system. In order to test the validity
of our variational approach (VA), we perform exact numerical
calculations on small clusters by means of the Lanczos algorithm.
We improve the previous exact diagonalization (ED) analysis of the
model, investigating small clusters up to six sites.
\cite{massimo} As shown in Fig. 3, each variational and exact
numerical curve exhibits a kink with increasing the el-ph
coupling. We have checked that at these couplings the effective
next-nearest-neighbor hopping changes sign opening an unphysical
region of the parameters. The agreement between numerical data and
variational approach is very good up to g values close to the
unphysical transition.

In order to characterize the polaron formation we also analyze the
electron-lattice correlation function $\chi_{i,\delta}$ defined in
Sec. III. In particular in Fig. 4 we show the behavior of
$\chi_{i,\delta}$ as a function of the SSH coupling for $\delta
=0,1,2$ and $\tilde t =2.5$. As expected, variational results and
exact numerical data always recover the perturbative values in the
limit of small el-ph coupling. Increasing $g$ the monotonic
behavior of the correlation function exhibits a kink, as the
ground state energy. In particular the correlation function at
next-nearest-neighbor ($\delta= 2$) changes sign as the effective
hopping, confirming  the pathological behavior. At couplings where
the ground state energy and the correlation
function show the kink, also the average phonon number is characterized by an
anomalous behavior as shown in the bottom right panel of Fig. 4.

In order to extract information on the values of $g$ at which
polaron crossover begins, before the opening of the unphysical
region, we also investigate the behavior of the quasipartcle
spectral weight $Z(0)$. We find that increasing the el-ph coupling
for fixed values of $\tilde t$, the spectral weight starts to drop
but it never reaches a really small value before the unphysical
sign change of the hopping occurs. Nevertheless we observe
distinct signatures of the tendency towards localization, as shown
in Fig. 5, where $Z(0)$ is plotted as a function of $g$ for the
fixed value $\tilde t=2.5$. 

We conclude our analysis collecting
the obtained data in the phase diagram of Fig. 6. 
It is calculated from the position of the kink in the ground
state energy obtained by means of  the variational approach (diamonds) and 
the exact diagonalization (triangles). The agreement
between the two methods becomes better moving towards the adiabatic limit. In
analogy with the phase diagram obtained for the Holstein polaron,
\cite{cataud1} we also mark a crossover region defined as the
range of parameters for which $Z(0)$ is less than 0.9. As shown in
Fig. 6, we find that the considered SSH model does not present any
marked mixing of electronic and phononic degrees of freedom, being
the strongly coupled state prevented from the pathology of the
model. As far as the fully adiabatic limit $\omega_0 =0$ is concerned, 
we verify that the crossover line joins onto the line for the transition to the
unphysical region at the critical value $\lambda=0.25$, confirming
the discussion in Ref. [\onlinecite{massimo}].
We finally notice that, as discussed in Ref. [\onlinecite{massimo}], both
the crossover region boundary, and the instability line obtained by 
exact diagonalization are 
only weakly dependent on the adiabatic ratio, and that $\lambda$ is the
relevant electron-phonon coupling regardless the value of $\tilde{t}$.
This is a peculiarity of the SSH coupling with respect to the Holstein
one, where the polaron crossover moves to large values of $\lambda$ as 
the phonon frequency increases. \cite{massimo,ciuchi,ccg,caponeciuk}

\section{conclusions}
In this work we discussed the features of one electron non-locally
interacting with optical phonons in a discrete chain. We
introduced a variational wave function to locate the crossover
region for the transition between weak and strong localized
polaron solutions. In particular we found that the pathological
sign change of the effective next-nearest-neighbor hopping always
precedes a stable strongly localized solution. Such an unphysical
region of the model parameters does not occur in the case of
acoustical phonons being the deformation linked to the particle
extension along the entire chain. \cite{angilella} 
However we have also shown that, for finite values of the adiabaticity 
parameter, when
the phonon frequency is not really small, the non local (SSH)
el-ph interaction is more effective than the local (Holstein) one
in reducing the mobility of the electron.
Then our variational calculations are an interesting starting
point to examine the complex problem of the polaron formation in a
model where both local and non-local el-ph interactions are
present. In particular we emphasize that the proposed variational
wave function for the SSH limit can be slightly modified to be
suitable for the treatment of the complex case where both
interactions are present. Detailed future investigations in this
direction are required. Finally we stress that the validity of our
variational results is supported by an accurate analysis of exact
diagonalization data on small clusters. The agreement between VA
and ED data is good up to coupling values close to the unphysical
region.

\section{acknowledgements}

M. C. acknowledges the hospitality and financial support of the
Physics Department of the University of Rome "La Sapienza", as
well as the INFM, UdR Roma 1 and SMC, and Miur Cofin 2001.

\section*{Appendix A}
In the limit of small el-ph couplings, the perturbative second
order correction $\Delta E(k)$ to the tight-binding free band
energy is
\begin{eqnarray} \Delta E(k) &=& -4g^2\omega_0 \left[
\frac{1+2\tilde t \cos k}{4\tilde t^2} +
\frac{sink^2}{\sqrt{1+4\tilde t\cos k-4\tilde t^2(1-\cos^2 k)}}-
\frac{\sqrt{1+4\tilde t\cos k-4\tilde t^2(1-\cos^2 k)}}{4\tilde
t^2}\right] \nonumber \\
&&\nonumber \\
& & -g_1^2 \omega_0
\frac{1}{\sqrt{1+4\tilde t\cos k-4\tilde t^2(1-\cos^2 k)}}.
\label{e2}
\end{eqnarray}
Moreover, using the bare phonon and electronic Green propagators,
the perturbative self-energy reads
\begin{eqnarray}
\Sigma(k,\omega) &=& \frac{4g^2\omega_0}{N}\sum_q
\frac{[sin(k+q)-sink]^2}{\omega-\omega_0-\varepsilon(k+q)+i\delta}+
\frac{g_1^2\omega_0}{N}\sum_q
\frac{1}{\omega-\omega_0-\varepsilon(k+q)+i\delta}.
\label{realsigma}
\end{eqnarray}
From Eq. (\ref{realsigma}) we obtain the momentum dependent spectral weight
\begin{eqnarray}
Z(k)^{-1}
&=& 1-\frac{2\lambda}{\tilde t} \left[ 1 -
\frac{4 \tilde t^2 \sin^2 k (1+2\tilde t \cos k)}{
(1 +4\tilde t-4\tilde t^2 (1-\cos^2 k))^{ 3/2}} -
\frac{(1+2\tilde t \cos k)}{\sqrt
{1 +4\tilde t-4\tilde t^2 (1-\cos^2 k)}} \right] \nonumber \\
&& \nonumber \\
&+& 2 \lambda_1 \tilde t \frac{(1+2\tilde t \cos k)}{ (1 +4\tilde
t-4\tilde t^2 (1-\cos^2 k))^{ 3/2}}.
\end{eqnarray}

\section*{Appendix B}
In this appendix we report the matrix elements between the states
$|\psi_k^{(R)}>$ and $|\psi_k^{(L)}>$. These quantities are involved
in the calculation of the ground-state energy within the
variational approach.
We find
\begin{equation}
\langle \psi_k^{(L)} | \psi_k^{(R)} \rangle = \sum_{m_1, m_2}
\phi_k^{*(R)}(-m_1) \phi_k^{(R)}(m_2)Z_k^{(L-R)}(m_1-m_2),
\end{equation}
where the  phonon matrix element $Z_k^{(L-R)}(i-j)$ is defined as
\begin{equation}
Z_k^{(L-R)}(i-j)= _{ph}\langle0|
e^{-[U_k^{(L)}(j)+U_k^{(L)}(j-1)+U_k^{(L)}(j+1)]}
e^{-[U_k^{(R)}(i)+U_k^{(R)}(i-1)+U_k^{(R)}(i+1)]} |0\rangle_{ph}.
\end{equation}
Then we have
\begin{eqnarray}
&\langle& \psi_k^{(L)} |H_{kin}| \psi_k^{(R)} \rangle = -t
\sum_{m_1, m_2} \phi_k^{*(L)}(m_1)
\phi_k^{(R)}(m_2)[e^{ik}Z_k^{(L-R)}(m_1-m_2+1)
e^{-ik}Z_k^{(L-R)}(m_1-m_2-1)], \nonumber \\
& & \nonumber \\
&\langle& \psi_k^{(L)} |H_{ph} |\psi_k^{(R)} \rangle = -\omega_0
\sum_q \sum_{m_1,m_2} \phi_k^{*(L)}(m_1)
\phi_k^{(R)}(m_2)[w_q^*(k)]^2 Z_k^{(L-R)}(m_1-m_2)
e^{iq(m_1-m_2)},
\end{eqnarray}
and
\begin{equation}
\langle \psi_k^{(L)} |H_{int} |\psi_k^{(R)} \rangle = A_{1}+A_{2},
\end{equation}
with $A_1$ and $A_2$ given by
\begin{eqnarray}
&A_{1}&=\frac{g\omega_0}{\sqrt{N}} \sum_{q,m_1, m_2}
\phi_k^{*(L)}(m_1) \phi_k^{(R)}(m_2)w_q^*(k) e^{ik}
Z_k^{(L-R)}(m_1-m_2+1) \left[e^{iq(m_2-1)}(1-e^{iq})+e^{-iq
m_1}(e^{-iq}-1)
\right] \nonumber \\
& & \nonumber \\
&A_{2}& = \frac{g\omega_0}{\sqrt{N}} \sum_{q,m_1, m_2}
\phi_k^{*(L)}(m_1) \phi_k^{(R)}(m_2)w_q^*(k)
 e^{-ik}Z_k^{(L-R)}(m_1-m_2-1)\left[e^{iqm_2}(1-e^{iq})
+e^{-iq(m_1-1)}(e^{-iq}-1) \right].
\end{eqnarray}

The quantity $\varepsilon_k=\langle\psi_k^{(L)}
|H|\psi_k^{(L)}\rangle= \langle\psi_k^{(R)}
|H|\psi_k^{(R)}\rangle$ is easily derived using the matrix
elements given above.


\newpage
\begin{figure}[htb]
\begin{center}
\includegraphics[width=8cm]{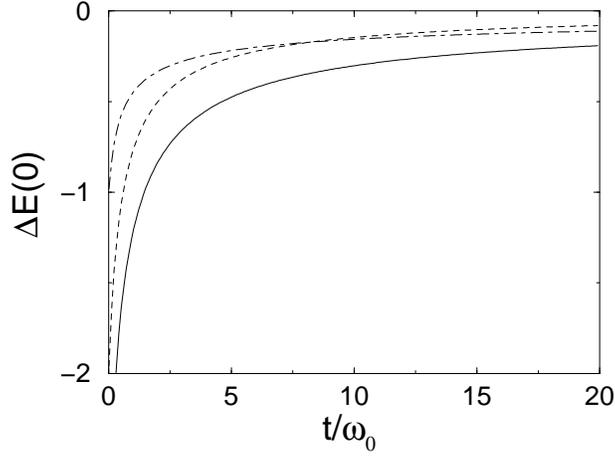}
\end{center}
\caption{Second order correction to the gound state energy, 
Eq. (\ref{egs}),
as a function of the adiabatic inverse ratio $\tilde t$, for $g=g_1=1$ (solid line). 
The dashed line is the SSH-like contribution ($g=1$ and $g_1=0$),
the dot-dashed line is the Holstein contribution ($g=0$ and $g_1=1$).
}
\label{epert}
\end{figure}
\vskip 0.5truecm

\begin{figure}[htb]
\begin{center}
\includegraphics[width=7cm]{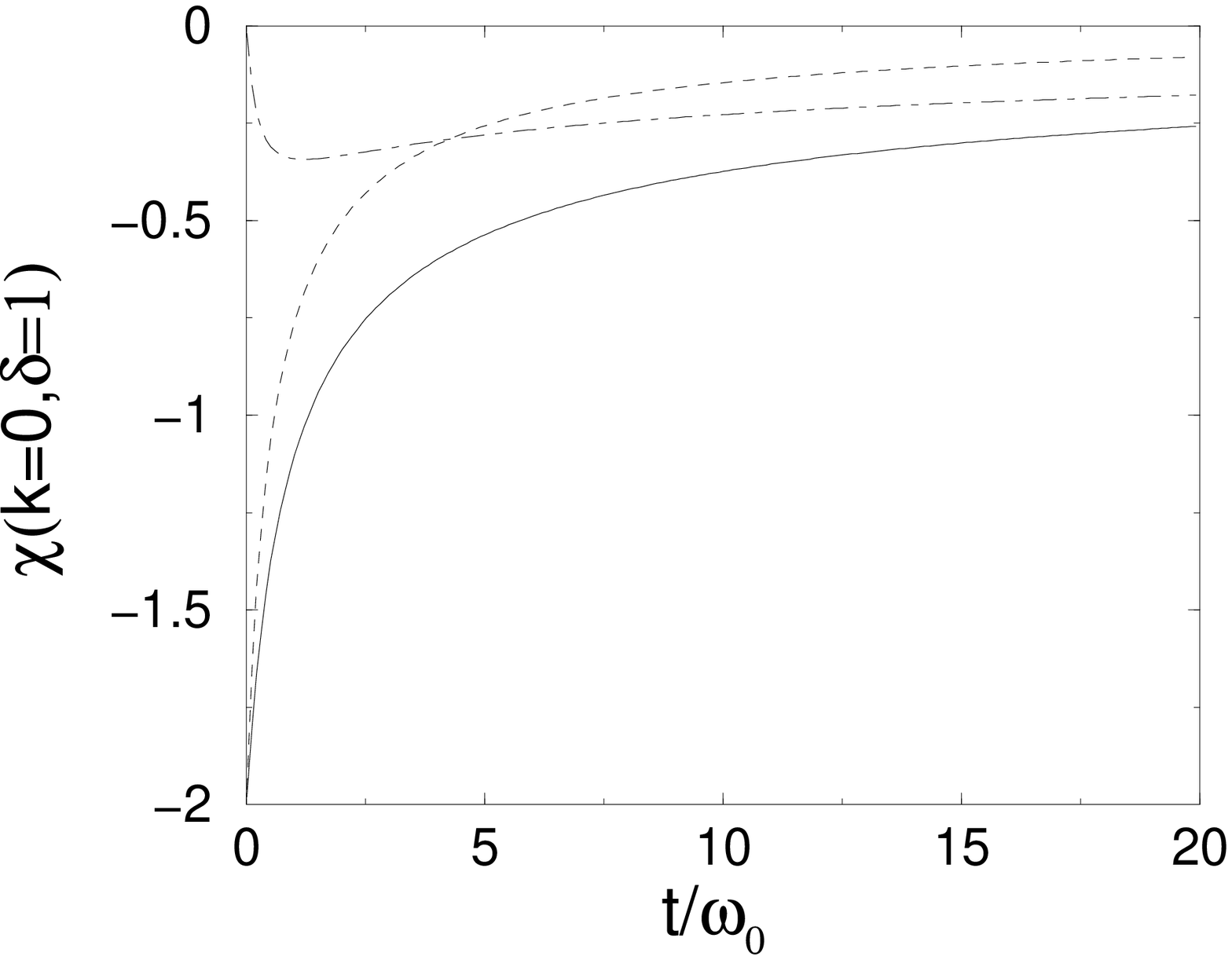}
\hskip 0.5truecm
\includegraphics[width=7cm]{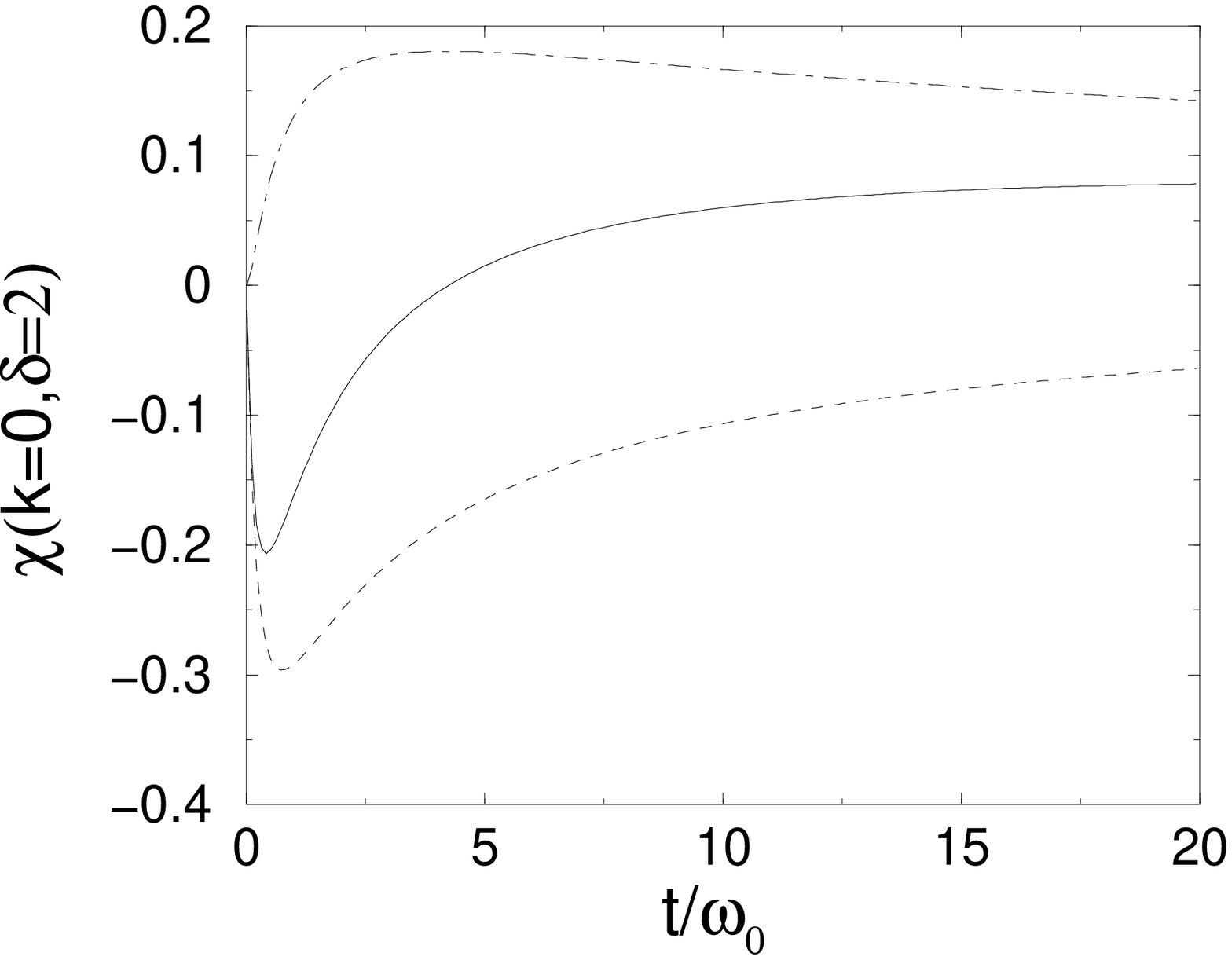}
\end{center}
\caption{Left: Correlation functions at nearest-neighbor sites (on the
left) and next-nearest-neighbor sites (on the right) at $k=0$ as
functions of the inverse adiabatic ratio $\tilde t$, for $g=g_1=1$ (solid line).
The dashed lines show the SSH-like contribution ($g=1$ and $g_1=0$),
the dot-dashed lines  the Holstein ones ($g=0$ and $g_1=1$).}
\label{corrpert}
\end{figure}
\vskip 0.5truecm

\begin{figure}[htb]
\begin{center}
\includegraphics[width=8cm]{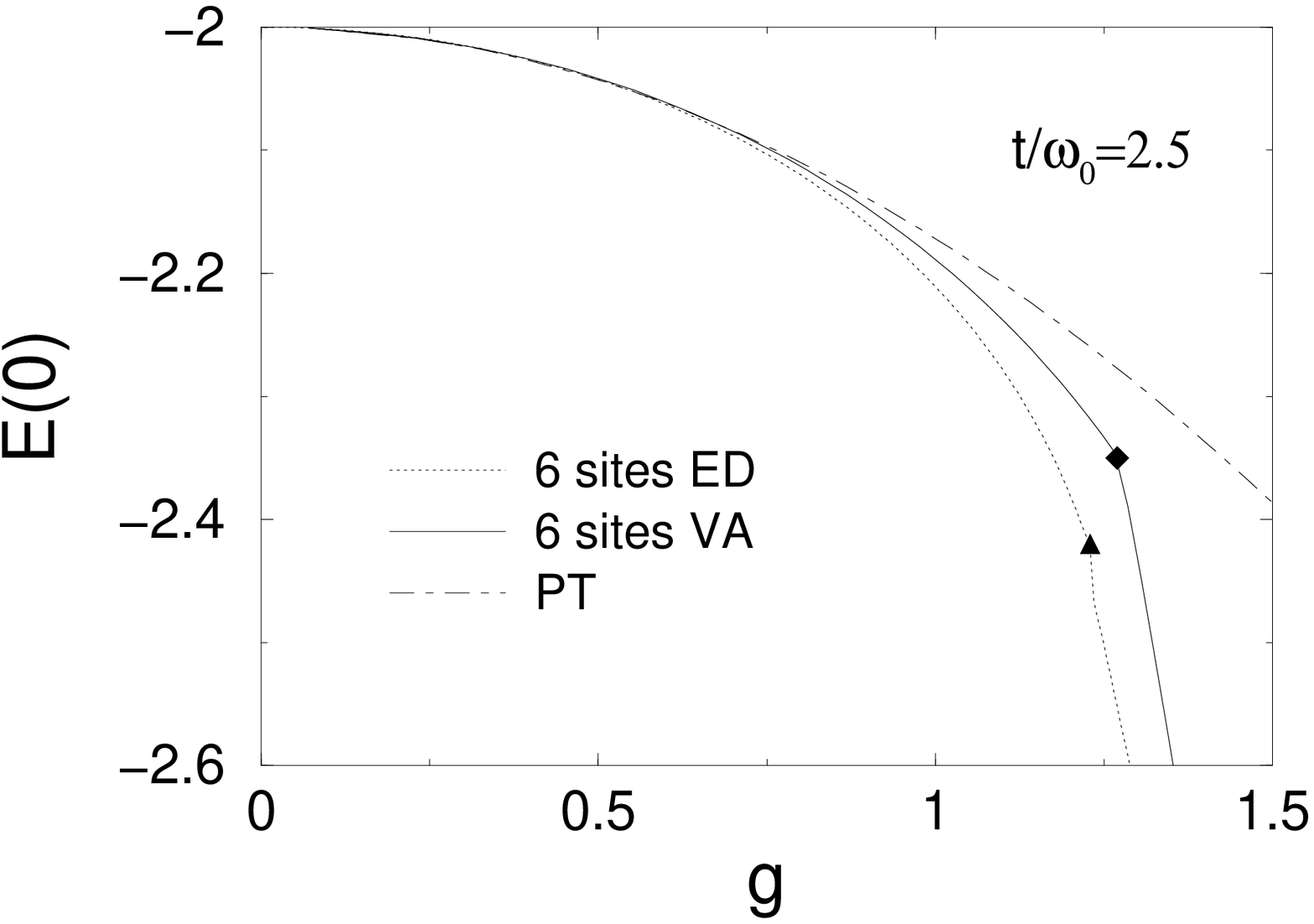}
\hskip 0.5truecm
\includegraphics[width=8cm]{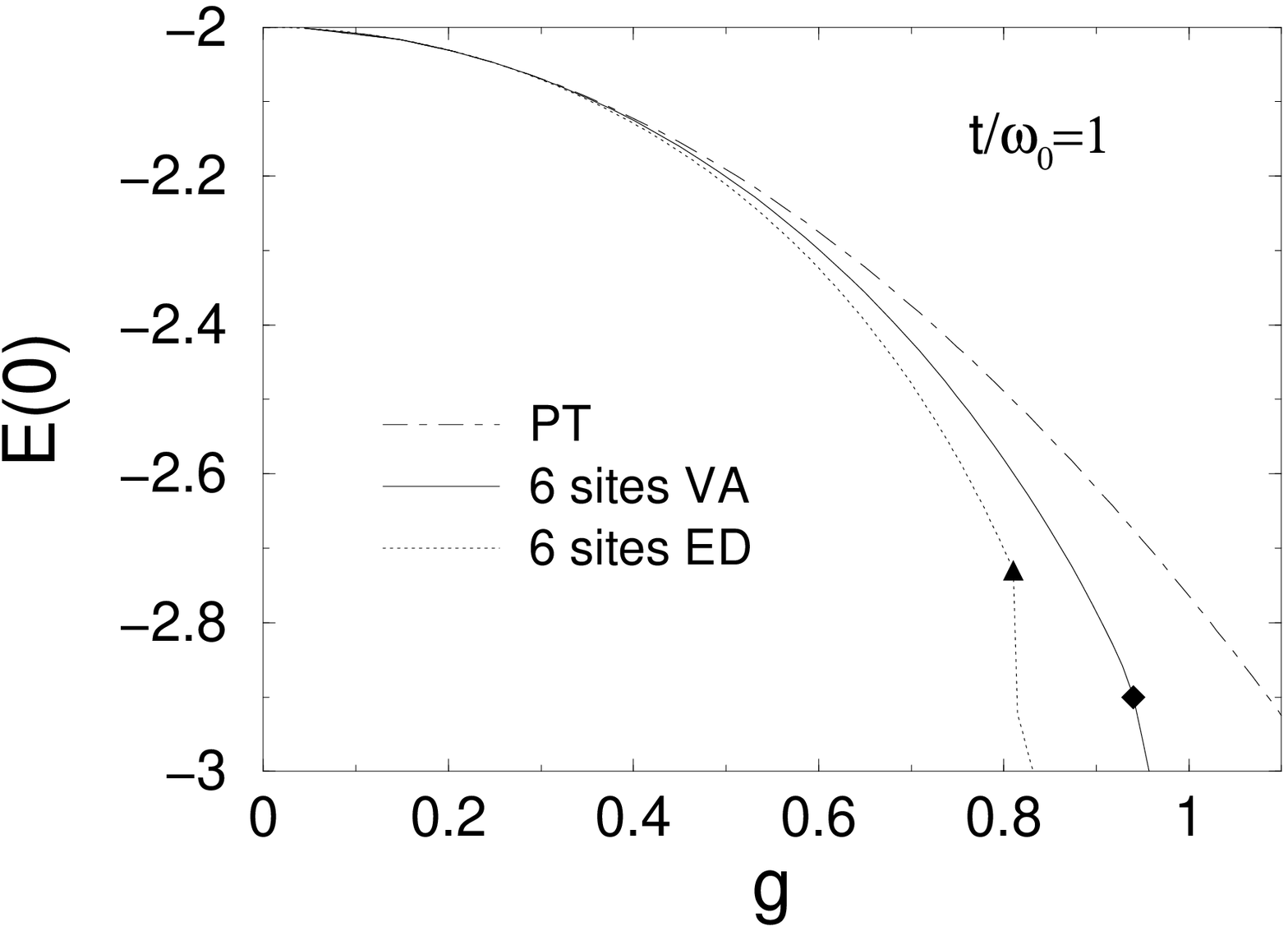}
\end{center}
\caption{Ground state energy $E(0)$ as a function of the SSH el-ph coupling $g$ for 
two different values of the inverse adiabatic ratio $\tilde t=2.5$ (left) and $\tilde t =1$ (right).
Solid and dotted lines 
are obtained from the variational approach and 
the Lanczos data for a six-site lattice, respectively; 
perturbative curves (dot-dashed lines) are plotted for comparison.
Symbols mark the kink values of the energy.}
\label{eva}
\end{figure}
\vskip 0.5truecm

\begin{figure}[htb]
\begin{center}
\includegraphics[width=12cm]{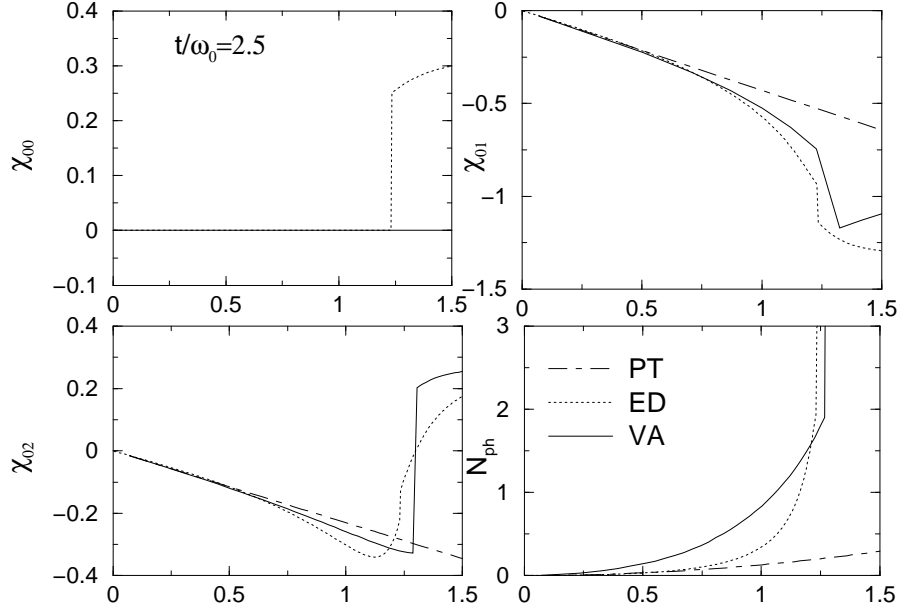}
\end{center}
\caption{Correlation functions $\chi_{k=0,\delta}$ with $\delta =0$ (top left),
$\delta =1$ (top right) and $\delta =2$ (bottom left) as functions of the
SSH coupling $g$ for $\tilde t =2.5$.  Bottom right: Phonon number vs. g 
for the same value of $\tilde t$.
Solid lines are obtained from the variational approach in the thermodynamical limit; dotted lines
show Lanczos data; perturbative curves from Eqs. (\ref{F0}) and Eq. (\ref{nph0}) with $g_1 =0$
(dot-dashed lines) are plotted for comparison.}
\label{corrva}
\end{figure}
\vskip 0.5truecm

\vskip 0.5truecm
\begin{figure}[htb]
\begin{center}
\includegraphics[width=8cm]{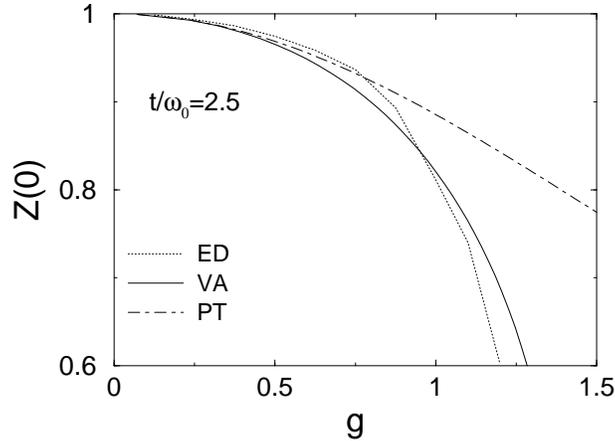}
\end{center}
\caption{Spectral weight $Z(0)$ as a function of the SSH el-ph coupling $g$ 
for $\tilde t =2.5$.
The solid line is obtained from 
the variational approach in the thermodynamical limit; the dotted line
shows Lanczos data; the perturbative curve from Eq. (\ref{zk0}) with $g_1 =0$  
(dot-dashed line) is plotted for comparison.}
\label{spec2}
\end{figure}
\vskip 0.5truecm

\vskip 0.5truecm
\begin{figure}[htb]
\begin{center}
\includegraphics[width=8cm]{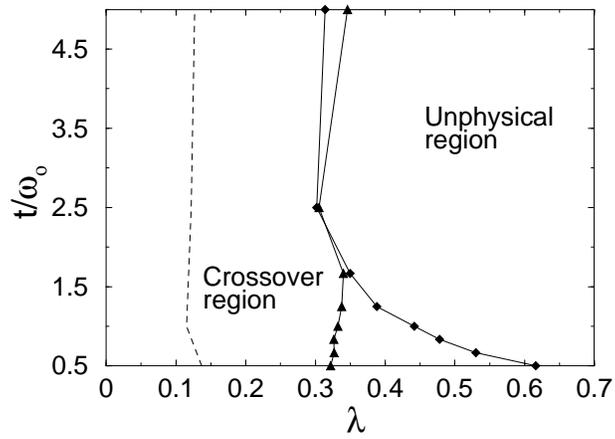}
\end{center}
\caption{Phase diagram for one electron in a six-site lattice.
Triangles and diamonds correspond, respectively, to the couplings where the 
exact numerical ground state energy and the variational result have a kink.
The dashed line indicates the boundary of the crossover 
region, where the spectral weight $Z(0)$ is less than $0.9$.}
\label{phd}
\end{figure}
\vskip 0.5truecm

\end{document}